
\magnification=1200

\nopagenumbers
\null
\vglue .7truein
\centerline{\bf DO COSMIC STRINGS}
\vskip .3truein
\centerline{\bf GIVE RISE TO VACUUM FLUCTUATIONS ?}
\vskip.7truein
\centerline {A. H. Bilge$^*$}
\vskip .2cm
\centerline {Department of Mathematics,}
\centerline{TUBITAK Marmara Research Center, Gebze Kocaeli, Turkey}
\vskip .5cm
\centerline{ M. Horta\c csu }
\vskip .2cm
\centerline{Physics Department,}
\centerline{TUBITAK Marmara Research Center, Gebze Kocaeli, Turkey}
\centerline{I.T.U., 80626, Maslak,Istanbul,Turkey}
\vskip .5cm
\centerline{ N. \"Ozdemir}
\vskip .2cm
\centerline{Physics Department,}
\centerline{I.T.U., 80626, Maslak,Istanbul,Turkey}
\vskip 4cm
\centerline{\bf Abstract}
The role of the identification of the vacuum and non-vacuum space-times
in the computation of vacuum fluctuations in the
presence of a cosmic string is discussed and an alternative interpretation of
the renormalization is proposed. This procedure does not give rise to vacuum
fluctuations.

\vfill
\noindent
$^*$Present Address: Department of Mathematics, Anadolu
University, Eski\c sehir, Turkey

\vfill\eject
\baselineskip=18pt
\footline={\centerline{\folio}}
\pageno=1
\vglue .5cm

\noindent
{\bf 1. Introduction.}

Whether vacuum  fluctuations exist in the presence of cosmic strings
was an important  question   in recent years. This problem has been
studied by various authors and it was answered affirmatively
[1-5].   The method for the
computation of vacuum fluctuations is to obtain  the  Greens function in the
space-time with a cosmic string and to renormalize this  function by
subtracting the vacuum contribution, i.e. the
Greens function for the Minkowski space.  This renormalization process
involves the difference    of two functions  defined on different
manifolds, hence we need to define a map to identify the points on the
space-time with a cosmic string and the Minkowski space-time.
 In this paper we discuss the
the role of this  identification of in the renormalization process.

 In the
literature
the computation and the renormalization of the Greens function for the
space-time with
a cosmic string is adopted from [6]  where the Greens function for a
flat space with a wedge cut out is obtained and renormalized. We claim
that the identification scheme underlying this renormalization is not
appropriate for the identification of a space-time with a cosmic string
and the Minkowski space,
and we propose a different identification which leads
to a renormalization that do
not remove the singularity. We then, calculate the energy difference
treating
the energy shift, as if this is due to a perturbation in Minkowski space,
and show that the difference in energies
is proportional to the energy itself.  Hence we conclude that
subtracting the vacuum can not tame the singularity.

We recall that an infinitely long straight cosmic string may be described by
a locally flat space-time with a ``topological defect". That is in cylindrical
coordinates, the range of the polar angle is $(0,2\pi\beta)$ with $\beta<1$
instead of being $(0,2\pi)$, but otherwise the metric is flat. This description
of the cosmic string provides an analogy with the computation of the Greens
function for a flat space with a wedge cut out [6].
In fact the expression of the Greens function for the string
in the coordinate system above  can be obtained directly from
the Greens
function of the flat space with a wedge, by imposing appropriate boundary
conditions. However there is a crucial difference in the
renormalization as discussed in Section 3.
\vskip .5cm
\noindent
{\bf 2. Preliminaries.}

Let $M$ be the spacetime manifold with a metric $g$, and $(U,h)$, where $h$ is
a homeomorphism from $U$ to $M$ be a chart domain. In the computations below we
emphasize the distinction between the quantities on $M$ and their local
coordinate expressions. For example,
$g$ is the metric on the manifold, and $g\circ h$ is its expression
in local coordinates. Similarly
the Greens function $G(p,p')$, $p,
p'\in M$ is a function on the manifold, but $(G\circ h)(x,x')$, $x,x'\in U$ is
a function on $U$. In practice we obtain $G\circ h$ by solving the Klein-Gordon
equation in $U$ with boundary conditions that reflect the physical situation on
the spacetime $M$.

We note that, here (and elsewhere in the literature except [7], where a
$C^\infty$ metric is used) the space-time with a cosmic string has a time
independent
metric, hence this metric does not represent the ``formation" of the cosmic
string. The information related to the formation of the cosmic string is coded
in the subtraction of the vacuum contribution.

We list below the Greens function used in the literature, with an emphasis on
the local coordinates used in each case. The computation methods are standard
and are given in [8]. We use the metric signature $(+,-,-,-)$.

\noindent
{\it Minkowski space with cylindrical coordinates:}
Let $M_o$ be $R^4$ with flat metric $g_o$.
The usual
cylindrical
coordinates provide a (global) coordinate chart. The chart domain $U$ is
$$U=\{(t,r,z,\phi)\in R^4:
-\infty< t< \infty,
             0< r<\infty,
             -\infty< z<\infty,
             0< \phi< 2\pi\}\eqno(2.1)$$
and
$g\circ h$ leads to the line element
$$ds^2=dt^2-dr^2-dz^2-r^2d\phi^2.\eqno(2.2)$$
In these coordinates the Greens function  $G_o\circ h$ can be obtained as the
solution of the Klein-Gordon equation with the boundary condition  $(G\circ
h)\mid_{\phi=0}=(G\circ h)\mid_{\phi=2\pi}$.
If we identify all the coordinates except $\phi $ and $\phi'$,
$G\circ h$ reduces to
$$(G\circ h)(\phi,\phi')={1\over
{16 \pi^2 r^2  \sin^2{(\phi-\phi')\over 2}}}\eqno(2.3)$$
This is the standart result for vacuum.
\vskip .2 cm

\noindent
{\it Minkowski space with a wedge cut out:} Let $N$ be the subset of
the Minkowski space with a wedge cut out. The metric $g'$ is the restriction of
 Minkowski metric to this subset. We use the same (global) coordinate
chart as above with the restriction of the domain. Namely
$$V=\{(t,r,z,\theta)\in R^4:
-\infty< t< \infty,
             r_0< r<\infty,
             -\infty< z<\infty,
             0< \theta< 2\pi\beta,\quad \beta<1\}\eqno(2.4)$$
and $h'=h\mid_{V}$. In this case the Klein-Gordon equation is same as
above but boundary conditions for the Greens
are different.
As a special case, if we impose periodic
boundary conditions, i.e. $(G\circ h')\mid_{\theta=0}=(C\circ
h')\mid_{\theta=2\pi\beta}$,
 and take the coincidence limit for all other coordinates, we obtain
$$ (G\circ h')(\theta,\theta)={1\over{16 r^2 \beta^2 \pi^2
\sin^2{(\theta-\theta')\over 2\beta} }}. \eqno(2.5)$$
This is the result in [6].

\vskip .2cm
\noindent
{\it  Space-time with cosmic string:} We recall that a spacetime $M$ with an
infinitely long cosmic string along the $z$ axis is locally a flat spacetime
with a conical defect.
There are two standard local coordinate
descriptions for this space-time: One can either ``straighten" the conical
surface to a plane by cutting out a wedge, or  ``stretch" the conical
spacetime to map the surface of the cone to a plane.
In both cases cylindrical provide a (global) chart.

\noindent
(i) ``Straighened"  coordinates $(V,k)$: $V$ is the open subset of $R^4$
described
above in (2.4) and $k$ is a homeomorphism of  $V$ into  $M$.
The expression of the metric in these coordinates, i.e.
$g\circ k$ leads to the line element
$$ds^2=dt^2-dr^2-dz^2-r^2d\theta^2\eqno(2.6)$$
In this coordinate system, the local expression of the Klein-Gordon operator is
the same as in the Minkowski space, but the
boundary conditions are
$(G\circ k)\mid_{\theta=0}=(G\circ k)\mid_{\theta=2\pi\beta}$. Hence the
expression
of the Greens function in these coordinates, i.e. $G\circ k$ is given by
(2.5), which is the result given in [1].

\noindent
(ii) ``Stretched" coordinates $(U,l)$: $U$ is the open subset of $R^4$
described in (2.1),
and $l$ is a homeomorphism of $U$ into $M$.
The chart transformation between these two
coodinates  is given by $\{t,z,r,\theta\}\to\{t,r,z,\phi\}$ where
$\phi=\theta/\beta$
The expression of the metric in these coordinates,
i.e. $g\circ l$ leads to the line element
$$ds^2=dt^2-dr^2-dz^2-\beta^2r^2d\phi^2,\quad\quad \beta<1.\eqno(2.7)$$
The expression of the Greens function in these coordinates after the
coincidence limit is then
$$(G\circ l)(\phi,\phi')={1\over{16 \pi^2 r^2 \beta^2 \sin^2{(\phi-\phi')\over
2}}}.\eqno(2.8)$$
This result coincides with that of [1] if we set $\phi=\theta/\beta$

Up the this point there is no problem in working with different coordinate
systems, since the Greens functions are related via coordinate transformations.
The problems arise at the stage of normalization where we need to identify the
points on the spacetime $M$ with the vacuum spacetime, and at this stage we
need to be careful in using local coordinates.

\vskip .5cm
\noindent
{\bf 3. Renormalization.} The description of the vacuum contribution is the
subtle part of the problem. The Minkowski space and the space-time with a
cosmic
string are two distint manifolds, and $G$ and $G_o$ are functions on these.
In order to subtract the vacuum contribution, we need to define a
map identifying the points on $M$ and $M_o$.  Let
$\varphi:M\to M_o$
be this identifying map (see Figure 1). Then the renormalized Green function
$G_r$ can be defined as a function on the manifold $M$, as
$$G_r(p,p')=G(p,p')-G_o \big(\varphi(p),\varphi(p')\big).\eqno(3.1)$$

We claim that a point $p$ in $M$ with ``stretched" coordinates
$\{t,r,z,\phi\}$  should be identified with the point $p_o$ in the Minkowski
space $M_o$ with coordinates $\{t,r,z,\phi\}$. Thus the
identifying map is
$$\varphi:M\to M_o, \quad \varphi=h\circ
l^{-1}\eqno(3.2)$$
where $h:U\to M_o$ and $l:U\to M$ are the coordinate functions described
above.
We can express the
renormalized Green function on  $U$: if $p=l(x)$, $x\in U$,
and $\varphi(p)=h(x)$, we have
$$G_r\circ l=G\circ l-G_o\circ h\eqno(3.3)$$
The expression of $G\circ l$ and $G_o\circ h$ are given respectively in (2.8)
and (2.3), hence the renormalization leads to the result
$$(G_r\circ l)(\phi,\phi')={1\over{16 \pi^2 r^2  \sin^2{(\phi-\phi')\over
2}}}\left({1\over \beta^2}-1\right).\eqno(3.4)$$
This result shows that the renormalization does not eliminate the singularity.

In the wedge calculation, one should identify a point $p$ in the flat space
with a wedge cut out, with coordinates $\{t,r,z,\theta\}$ with the point in the
Minkowski space with {\it same} coordinates, in other words, the
identification map $N\to M_o$ is the inclusion.
Then the
renormalized Green function is
$$(G_r\circ h')=(G\circ h')-(G_o\circ h\circ i)$$
where $i$ is the inclusion map from $V$ to $U$. Then
the renormalization gives a finite result as obtained in [6].

\noindent
{\bf 4. The perturbation approach.}
An alternative computation  of this phenomena may be instructive
 at this stage to justify our claim even more strongly.
Essentially we have a scalar field $\Phi$ in a flat space with a defect.  We
can
rewrite the problem as a scalar particle subject to an interaction where the
Hamiltonian density is given as

$$H=H_0+ V\eqno(4.1) $$
$$ H_o= {\partial_{\mu} \Phi }{\partial ^{\mu} \Phi }\eqno(4.2) $$
$$V= \left( {1\over \beta^2}-1 \right) \left( {\partial\Phi\over
\partial \phi} \right)^2\eqno(4.3)$$
This is an exactly solvable model.  Assuming exponential behaviour for time
dependence, we can calculate the energy eigenvalue $\omega$ in both the
perturbed and
unperturbed cases.  If we calculate the total energy, we see that
$$ \int d^3p\  \omega_\beta  \propto \Lambda^4 \eqno(4.4)$$
$$ \int d^3p\  \omega \propto \Lambda^4
\eqno(4.5)$$
$$\int d^3p\  (\omega_\beta- \omega) \propto  \Lambda^4 \eqno(4.6)$$
Here $\omega_{\beta} $ and $\omega$ are the energy eigenvalues with and
without the interaction and $\Lambda$ is the cut-off.  If the subtraction of
the
vacuum resulted in a regularization, the difference of the energies would
have a milder divergence.

One consistent way to make this contribution finite is to use a counterterm
that cancels it completely.  Since the
interaction is only bilinear, we can not regenerate it in the next order,
as in the case with trilinear or higher couplings.
  We get zero for the energy with this regularization.

\vskip .5cm
\noindent
{\bf 5. Conclusion.}

Here we tried to point to the difference in identifying the vacuum in the case
of a wedge calculation in electromagnetic theory and in a spacetime with cosmic
string.
These two situations can be described by with the same metric hence the local
coordinate formulation of the Greens function is  the same.
However the renormalization process, i.e. the
quantum part of the problem
involves the identification of the the points in different manifolds. Our claim
is that the physically meaningful identification corresponds to subtracting
(2.3) from (2.8) which results in (3.4).
Since we cannot tame the singularity by subtracting the vacuum, we suggest that
this whole term should be subtracted, thus yielding no vacuum fluctuations in
the presence of a cosmic string.

\bigskip
\noindent
{\bf Acknowledgements.}
We are grateful to Prof. Yavuz Nutku for suggesting this problem to us.
We  acknowledge very
helpful discussions with Profs. \.I.H.Duru, \" O. F. Day{\i} and J.Kalayc{\i}.
This work is partially supported by T\"UBITAK, The Scientific and
Technical Research Council of Turkey under TBAG \c CG-1. \bigskip
\vfill
\eject
\vglue 1cm
\baselineskip 14pt
\noindent
{\bf REFERENCES}
\vskip .5cm
\item{1.}  T.M.Helliwell and K.A. Konkowski,  Phys.Rev. D  34, 1918
(1986).
\item{2.}  A.G.Smith, Tufts preprint (1986), published in
``The formation and Evolution of Cosmic Strings", ed. by G.Gibbons,
S.Hawking and T.Vachaspati, Cambridge Univ. Press (1990) p.263-292.
\item {3.}  J.S.Dowker, p. 251-261 in the above cited book.
\item {4.}  B.Linet ,Phys.Rev. D 35, 536 (1987).
\item {5.}  V.P.Frolov and E.M. Serebriany, Phys.Rev. D34, 3779 (1987)
\item {6.}  D.Deutsch and P. Candelas, Phys. Rev. D 20, 3063 (1979).
\item {7.}V. Sahni, Modern Phys. Lett.A3, 1425 (1988).
\item{8.} N.D. Birrell and P.C.W. Davies, ``Quantum Fields in Curved Space",
(Cambridge University Press) 1982.

\end

We can show the existence of an ambiguity in the renormalization scheme
for this problem by using still another form of the same metric, given by
Gleiser and Pullin $^{/8}$.  These authors propose the metric

$$ ds^{2} = 4 du dv - (u-\beta^{2} v)^{3} - (u+\beta^{2} v)^2 dz^2 \eqno{19} $$
Here $\beta$ 's a constant less than unity, $\leq \phi \leq 2\pi $, $z$ has the
dimensions of
an angle but extends from minus infinity to plus infinity. The
 authors $^{8}$  transform this metric to
 $$ ds^{2} = dT^{2} -dR^{2} -\beta^{2} R^{2} d\Phi^{2} - dZ^{2} \eqno{20} $$
 by defining
 $2\beta v=(T^{2}-Z^{2})^{1/2} -R$, $2u \beta^{-1} v = (T^{2}-Z^{2})^{1/2}+R$,
 $\phi = \Phi $ ,$z ={1\over \beta} tanh ^{-1} ({Z\over T}) $.
 This is the same metric H-K starts, their eq. (2).  If one defines $ \beta
\Phi
= \Theta $
 their metric reduces to
 $$ ds^{2} = dT^{2}- dR^{2} - R^{2} d\Theta^{2} -dZ^{2} \eqno {21} $$

 We do not have to stick to this form of the metric in a classical
 calculation.  We use yet another transformation, on the original
Gleiser-Pullin
 metric, eq. (19).
 Any problems will only demonstrate that anomalies may arise when we take the
 coincidence limit, to get the expectation value of the field theoretical
 stress-energy tensor.

 We use $ u= {{r+t}\over 2}$, $ v={{t-r}\over{2\beta^{2}}}$, $x=
rcos{\beta\phi}
$,
 $y=rsin{\beta \phi}$ where $0\leq \phi \leq 2\pi $, and all the other
variables
 range from
 minus to plus infinity.  Then the metric reads
 $$ds^2={1\over{\beta^2}}(dt^2--dx^2-dy^2)-t^2 dz^2 \eqno{22} $$.
 As in eq.(19), $z$ acts as the "angular variable " here.  It is the
streographi
c
 projection on the sphere.

 Using standard methods we find that in the limit $x=x' $ and $y=y'$ are
identif
ied
 the Greens' Function, for this metric reads
$$ G ={1\over{16 \pi^{2} \beta (t+t')(t-t')}}
{\left( {2i\over{\beta}}ln{t'\over {t}}
\right)\over{-\left( ln{t'/t}\over{\beta} \right)^{2} -(z-z')^{2}} } \eqno{23}
$
$
Note that whereas one can expand the logarithm in a power series in $t'/t$,
$z-z
'$ is in the form of only a monomial.

At this point we have two alternatives in the coincidence limit.
We either take $t=t'$ first, and leave the identification of the "angular
variable" to the last.  Then
$$ G={i\over{16 \pi^{2} \beta^{2} t^{2} (z-z')^2}}    \eqno{24} $$
Note that there is no further term in this expression.
To make this expression finite at the
 coincidence limit, either we subtract this term totally, and obtain zero, or
su
btract only the
 vacuum,
$$ G_0 = {i\over{16 \pi^{2} t^2(z-z')^2}} \eqno {25} $$
and have a divergent result ; the same result we obtain when we subtract
eq. (17) from eq.(15).

Another choice would be to take $z=z'$ first.  Then $\beta $ disappears from
the
 expression totally.
Although we can get a finite expression both for the Greens' Function in the
coi
ncidence limit and the vacuum
expectation value of the stress-energy tensor , to achieve this result we are
no
t subtracting the vacuum.
If we define the vacuum Greens' Function to be of the same form as the original
expression
except with $\beta=1$, we will get the null result upon subtraction.
We think this example illustrates the ambiguity in the renormalization
precedure
for this metruc further.

\vskip .5 cm
\noindent
{\bf 4. Computation of the Greens function.}
We use the local coordinates $(U,l)$ hence start with the metric (2.2)
Hereafter we drop the explicit dependence on local coordiantes, and $G$ denotes
in fact $G\circ l$. For this calculation we use similar methods as given in
[2].

If one couples a massless scalar field to this metric, we
get
$$ \left[ {\partial^2\over{\partial t^2}}-{1\over r} {\partial \over {\partial
r
}} \left( r {\partial \over {\partial r}} \right)
-{1\over{r^2{\beta^2}}}{\partial ^2\over{\partial \phi ^2}}
-{\partial^2 \over\partial z^2} \right]  G = -\delta (x,x') \eqno(4.1)$$
The eigenfunctions of this operator are
$$ u(x;\omega,k,p,n)=\left( p\over \beta\right)^{1/2} {1\over(2\pi)^3}
e^{-i(kz-\omega t)} e^{-i \phi n} J _{n/\beta} (pr) \eqno(4.2)$$
Here $J_{\nu}$ is a Bessel function of first kind, $n$ is an integer;
the corresponding eigenvalues are $ \omega ^{2} - k^{2} - p^{2} $, where
$\omega$,$k$ are arbitrary real numbers, and $p$ is real and positive.
The Greens Function is formed, using the Schwinger formalism as
$$\eqalignno{
G(x,x')= i {\int_{-\infty}^{\infty}{d \omega\over (2\pi)}}
&{\int_{-\infty}^{\infty}{ dk \over (2\pi)}}{\int_0 ^{\infty} dpp}
{\int_0 ^{\infty} ds}{e^{is(\omega ^2 -k^2 -p^2)}}
{e^{-i\omega(t-t')}} {e^{ik(z-z')}} {1\over{2 \pi \beta}}\cr
& \times \sum_{-\infty}^{\infty} J_{n\over \beta} (pr)
J _{n\over \beta} (pr') {e^{in(\phi - \phi')}}&(4.3)\cr}$$

We use
$$ {\int_{-\infty}^{\infty}} d\omega
{e^{is{\omega}^2-i\omega(t-t')}}
= \sqrt{\pi\over{-is}} {\exp{{-i(t-t')^2\over{4s}}}}\eqno(4.4) $$
$$ {\int_{-\infty}^{\infty}} dk {e^{-isk^2-ik(z-z')}}
= \sqrt{\pi\over{is}} {\exp{{i(z-z')^2\over {4s}}}}\eqno(4.5) $$
$$ {1\over{i\pi}}{\int_0 ^{\infty}} {ds\over s} {\exp{\left(-{i\over 4s}
\left[ (t-t')^2-(z-z')^2 \right] -isp^2 \right)}}= {2iK_0 \over
\pi} (-ip \rho)\eqno(4.6) $$
where $K$ is the modified Bessel function,
$$\rho = \sqrt { (t-t')^2-(z-z')^2}\eqno(4.7)$$
and
$$ {\lim_{\epsilon\rightarrow 0} {\int_{0 ^{\infty}}}
dp p K_0(-i\rho p + \epsilon) J _{n \over\beta}(pr) J_{n \over\beta}(pr')
 = r_1^{-1} r_2^{-1} \Delta ^{n\over\beta}}  \eqno(4.8)$$
where
$$r_1=\left[ -(t-t')^2+(z-z')^2+(r-r')^2 \right]^{1/2} \eqno(4.9)$$
$$r_2=\left[-(t-t')^2+(z-z')^2+(r+r')^2 \right]^{1/2}\eqno(4.10)  $$
$$\Delta = \left( {{r_2-r_1}\over{r_2+r_1}}\right)\eqno(4.11)$$
We express the Greens Function as
$$ G(x,x')= {1\over{(2\pi)^2\beta}}{\sum_{n=0}^{\infty}}
{{\exp{in(\phi-\phi')} \Delta^{n\over\beta}}\over{r_1 r_2}} \eqno(4.12)$$
Here
$\Delta \leq 1 $
and equality is attained at the coincidence limit; so, we can perform the
summation
$$G(x,x')={1\over{(2\pi)^2\beta r_1 r_2}} \left( {1-\Delta ^{2\over\beta}
\over{1+\Delta^{2\over\beta}-2
\Delta^{1\over\beta}\cos(\phi-\phi')}}\right) \eqno(4.13)$$

At this point we can put $t=t'$,$z=z'$,$r=r'$ with impunity and get
$$G(\phi,\phi')={1\over{16 \pi^2 r^2 \beta^2 \sin^2{(\phi-\phi')\over 2}}}
\eqno(4.14)$$

Note that in the coordinate system $(V,k)$ one would obtain
$$ G(\theta,\theta)={1\over{16 r^2 \beta^2 \pi^2 \sin^2{(\theta-\theta')\over
2\beta} }} \eqno(4.15)$$
\medskip
{\bf Remark.} It can be seen that if we compute the coincidence limit by
setting $\phi=\phi'$ first in (4.13), the vacuum contribution, i.e. the
divergent part in the limit is independent of $\beta$, hence taking the limit
in this order leads to the same result as in [1-6]. Hence the result of the
renormalization depends on the order of taking the limits.
\medskip

Note that the
expression of the Greens function given by (4.13)
involves no approximations. This  expression represents a smooth function of
$(x-x')$ except at the origin, i.e. when $t=t'$, $z=z'$, $r=r'$, $\phi=\phi'$
and the divergence at the origin is proportional to $(r-r')^{-2}$. The above
remark shows that the  limit of the function $(r-r')^2G$ at the origin depends
on the direction of approach to the origin, hence the limit does not exist.
Therefore, the definition of the vacuum contribution as the divergent part of
the Greens function at the coincidence limit is ambiguious in this coordinate
system. Consequently, as this limit depends on the coordinate systems it is not
well defined.

Actually it is possible to transform the metric for the space-time with a
cosmic string (2.7) to the form
$$ds^2={1\over \beta^2}(dt^2-dx^2-dy^2)-t^2dz^2\eqno(4.16)$$
and calculate the Greens function as
$$G=-{1\over 8 \pi^2\beta}{1\over (t+t')(t-t')}
  {a\over a^2+(z-z')^2}\eqno(4.17)$$
where
$$a={1\over \beta}\ln {t'\over t}.\eqno(4.18)$$
In this case if we compute the coincidence limit by setting $t=t'$ first $G$
reduces to
$$G={1\over 16\pi^2\beta^2t^2(z-z')^2}\eqno(4.19)$$
which is a monomial with no finite part. But if we set $z=z'$ first $G$ is a
logarithmic function which is independent of $\beta$. This computation also
shows the ambiguity in the renormalization.


\vskip .5cm

%
describe a " cosmic string ", we think our